\documentclass[letterpaper, 10 pt, journal, twoside]{ieeetran}  % Comment this line out if you need a4paper

\IEEEoverridecommandlockouts                              % This command is only needed if 
\usepackage[dvips]{graphics} % for pdf, bitmapped graphics files
\usepackage{epsfig} % for postscript graphics files
\usepackage[fleqn]{amsmath} % assumes amsmath package installed
\usepackage[psamsfonts]{amssymb}  % assumes amsmath package installed
\usepackage{theorem}
\usepackage{subfigure}

\def\proof{\noindent\hspace{2em}{\itshape Proof: }}
\def\QEDclosed{\mbox{\rule[0pt]{1.3ex}{1.3ex}}} % for a filled box
\def\QED{\QEDclosed} % default to closed

\newtheorem{definition}{Definition}

\newtheorem{theorem}{Theorem}

\newtheorem{remark}{Remark}
\newtheorem{corollary}{Corollary}

\newcommand{\sk}{\mathrm{sk}}

\newcommand{\R}{{\mathbb R}}
\renewcommand{\S}{{\mathbb S}}
\newcommand{\cS}{{\mathcal S}}
\newcommand{\G}{{\mathcal G}}
\newcommand{\V}{{\mathcal V}}
\newcommand{\E}{{\mathcal E}}
\newcommand{\K}{{\mathcal K}}

\newcommand{\N}{{\mathcal N}}
\newcommand{\T}{{\mathrm T}}

\title{Distributed Collision-Free Motion Coordination on a Sphere: A Conic Control Barrier Function Approach}

\author{Tatsuya Ibuki$^{1}$, Sean Wilson$^{2}$, Aaron D. Ames$^{3}$, and Magnus Egerstedt$^{2}$% <-this % stops a space
\thanks{*This work was supported in part by JSPS KAKENHI under Grant 18K13775 and in part by the US National Science Foundation through Grants 1531195 and 1932091.}% <-this % stops a space
\thanks{$^{1}$T. Ibuki is with Department of Electronics and Bioinformatics, School of Science and Technology, Meiji University, Kanagawa 214-8571, Japan {\tt\footnotesize ibuki@meiji.ac.jp}}
\thanks{$^{2}$S. Wilson and M. Egerstedt are with the School of Electrical and Computer Engineering, Georgia Institute of Technology, Atlanta, GA 30332, USA {\tt\footnotesize\{sean.t.wilson, magnus.egerstedt\}@ece.gatech.edu}}
\thanks{$^{3}$A. D. Ames is with the Department of Mechanical and Civil Engineering, California Institute of Technology, Pasadena, CA 91125, USA {\tt\footnotesize ames@caltech.edu}}%
}

\begin{document}
\pagestyle{empty}

\maketitle
\thispagestyle{empty}

%%%%%%%%%%%%%%%%%%%%%%%%%%%%%%%%%%%%%%%%%%%%%%%%%%%%%%%%%%%%%%%%%%%%%%%%%%%%%%%%
\begin{abstract}
This letter studies a distributed collision avoidance control problem for a group of rigid bodies on a sphere.
A rigid body network, consisting of multiple rigid bodies constrained to a spherical surface and an interconnection topology, is first formulated.
In this formulation, it is shown that motion coordination on a sphere is equivalent to attitude coordination on the 3-dimensional Special Orthogonal group.
Then, an angle-based control barrier function that can handle a geodesic distance constraint on a spherical surface is presented.
The proposed control barrier function is then extended to a relative motion case and applied to a collision avoidance problem for a rigid body network operating on a sphere.  
Each rigid body chooses its control input by solving a distributed optimization problem to achieve a nominal distributed motion coordination strategy while satisfying constraints for collision avoidance.
The proposed collision-free motion coordination law is validated via simulation.
\end{abstract}

% Keywords appear just beneath the abstract.

%%%%%%%%%%%%%%%%%%%%%%%%%%%%%%%%%%%%%%%%%%%%%%%%%%%%%%%%%%%%%%%%%%%%%%%%%%%%%%%%
\section{INTRODUCTION}
%Motion coordination of a group of networked multi-robots is expected as an efficient technique to achieve better task performance and more robustness than single robot operation \cite{book:RN,book:egerstedt}.
%Distributed cooperative control is a powerful tool for such coordination, e.g., consensus, formation, flocking, and coverage control \cite{book:RN}-\cite{martinez_07CSM}.
\IEEEPARstart{S}{afe} and distributed motion coordination of individual robots within a multi-robot collective is required to solve many tasks, like formation, flocking, and coverage control \cite{book:RN}-\cite{martinez_07CSM}.
While many studies focused on motion coordination of a multi-robot system consider 3-dimensional (3D) Euclidean space or a 2-dimensional (2D) plane as a workspace, a spherical surface is also often required \cite{muralidharan_19TAC}-\cite{saber_06CDC}.
This motion coordination on a sphere is motivated not only by theoretical interests but also by some industrial application such as planetary-scale motion coordination/localization in the space/ocean, and vehicle coordination on the surface with a large radius of curvature.
Moreover, spherical motion constraints can be considered for motion control of manipulators or attitude control of pan-tilt cameras, and recently, the constraints are also taken into consideration in dynamics of cooperative transportation of a payload with multiple flying vehicles \cite{lee_14CDC, wu_14CDC}.
%This letter also focuses on distributed motion coordination on a sphere.

%At the stage of the implementation, collision avoidance is an essential requirement for networked multi-robots to guarantee safe motion coordination operation.
No matter what motion constrained workspace a multi-robot system operates in, effective collision avoidance is an essential requirement to guarantee hardware safety.
A potential-based approach is one common collision avoidance strategy for multi-agent systems \cite{li_14TAC}, \cite{verginis_19CSL}-\cite{dimarogonas_06Automatica}.
% and based on constraints in optimization problems .
This technique introduces a nonnegative scalar function that increases as a robot approaches obstacles, like other robots or environmental hazards.
Then, collision-free motion coordination methods incorporate its negative gradient to guarantee a safe operating distance.
However, the potential function often needs to be infinite at the obstacle boundary, which causes overcaution about safety, i.e., less control performance. 

%On the other hand, a numerical optimization method with constraints for safe task operation is recently presented 
More recently, constraint-based optimization methods have been used to guarantee robot safety during operation \cite{ames_17TAC}-\cite{ames_19ECC}.
Here, a scalar function describing a safe set, called a \textit{control barrier function (CBF)}, is introduced, and the forward invariance of the dynamics within the safe set is guaranteed via the constraints derived by the CBF.
In this approach, the control input is given by solving an optimization problem to achieve a control task \textit{as much as possible} while guaranteeing the safety.
This technique is also applied to collision-free motion coordination problems for multi-agent systems as in \cite{ibuki_20RAL,wang_17TRO,panagou_16TAC}.
Most of the existing studies, however, consider collision avoidance problems with standard Euclidean distances, i.e., in 3D space or on a 2D plane. % \cite{ibuki_20RAL,wang_17TRO,panagou_16TAC}.
This work extends the CBF-based approach to a collision-free motion coordination method on a spherical surface, where the safety is defined with geodesic distances.

This letter first formulates a rigid network consisting of multiple rigid bodies with their motion dynamics constrained to a spherical surface and an interconnection topology.
We show that motion coordination on a sphere is analogous to attitude coordination on the 3D Special Orthogonal group: $SO(3)$.
Then, as a bridge to a collision avoidance problem on a sphere, we develop a CBF-based safe control technique on $SO(3)$.
This approach is first applied to a cone-type (conic) constraint satisfaction problem, and by extending it to a relative motion case, we propose a collision-free motion coordination law for a rigid body network on a spherical surface.
In the proposed method, each rigid body selects its control input by solving a distributed optimization problem to achieve a given motion coordination task as much as possible while guaranteeing collision avoidance.
The effectiveness of the proposed approach is demonstrated via simulation.

The main contributions of this letter are twofold:
First, we develop a new CBF to handle constraints on $SO(3)$ by extending the classical CBF methods for vector fields presented in \cite{ames_17TAC}-\cite{ames_19ECC}.
Here, we also provide an example of safe attitude control for a single rigid body with a conic constraint.
Secondly, we extend this kind of CBF to a relative motion case, and propose a novel distributed collision-free motion coordination method for a rigid body network on a spherical surface.

%%%%%%%%%%%%%%%%%%%%%%%%%%%%%%%%%%%%%%%%%%%%%%%%%%%%%%%%%%%%%%%%%%%%%%%%%%%%%%%%
\section{PROBLEM SETTINGS}\label{sec:rbm}
\subsection{Rigid Body Motion}
As a preliminary, the motion dynamics of multiple rigid bodies in general 3D space are first introduced.
Let us consider a set of $n$ rigid bodies.
Each rigid body $i\in\{1,\dots,n\}$ has a body fixed frame $\Sigma_i$ in a world frame $\Sigma_w$.
The position and attitude of rigid body $i$ in $\Sigma_w$ are represented by $(p_{wi},e^{\hat\xi_{wi}\theta_{wi}})\in SE(3)$.
%:=\R^3\times SO(3),~SO(3):=\{R\in\R^{3\times 3}\mid RR^\T=I_3,~\det R=1\}$.
Here, $e^{\hat\xi\theta}\in SO(3)$ is the exponential coordinate of the rotation matrix with the rotation axis $\xi\in\R^3$ $(\|\xi\|=1)$ and angle $\theta\in[-\pi,\pi)$ \cite{book:robomani}.
The operator $\wedge:\R^3\to so(3)$ gives $\hat ab=a\times b$ for any 3D vectors $a,b\in\R^3$, and $\vee: so(3)\to\R^3$ is its inverse operator. 
For the ease of representation, $\hat\xi_{wi}\theta_{wi}$ is written as $\hat\xi\theta_{wi}$ throughout this letter.
%($I_n\in\R^{n\times n}$ is the $n\times n$ identity matrix.) ($O_n\in\R^{n\times n}$ is the $n\times n$ zero matrix.)

The translational and rotational body velocity of rigid body $i$ relative to $\Sigma_w$ is denoted by $v_i\in\R^3$ and $\omega_i\in\R^3$, respectively.
Then, for each rigid body $i\in\{1,\dots,n\}$, we have the following \textit{rigid body motion} \cite{book:robomani}:
\begin{eqnarray}
\dot p_{wi}=e^{\hat\xi\theta_{wi}}v_i,~\dot e^{\hat\xi\theta_{wi}}=e^{\hat\xi\theta_{wi}}\hat\omega_i.  \label{eq:rbm}
\end{eqnarray}
%

%%%%%%%%%%%%%%%%%%%%%%%%%%%%%%%%%%%%%%%%%%%%%%%%%%%%%%%%%%%%%%%%%%%%%%%%%%%%%%%%
\subsection{Rigid Body Motion on a Sphere}\label{subsec:rbms}
\begin{figure}[t]
\centering
\includegraphics[width=.84\linewidth]{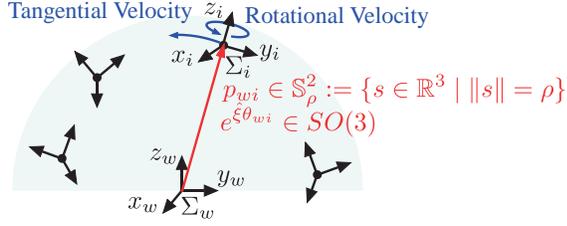}
\caption{Multiple rigid bodies on sphere. The origin of the world frame $\Sigma_w$ is located at the center of a sphere with the radius $\rho$, and the position of each rigid body is constrained on the surface of the sphere.}
\label{fig:rbm_sphere}
\end{figure}
Let us next consider the motion dynamics of multiple rigid bodies constrained to the surface of a sphere with the radius $\rho>0$.
Without loss of generality, let the origin of the world frame $\Sigma_w$ be located at the center of the sphere.
Suppose also that the direction of the $z$-axis of each body frame $\Sigma_i,~i\in\{1,\dots,n\}$ coincides with the radial direction  (see Fig. \ref{fig:rbm_sphere}).
In this case, each rigid body $i$ has the  spherical constraint with the basis axis $e_3:=[0~0~1]^\T\in\S^2:=\{s\in\R^3\mid \|s\|=1\}$ as
\begin{eqnarray}
p_{wi}=\rho e^{\hat\xi\theta_{wi}}e_3. \label{eq:s_constraint}
\end{eqnarray}

Substituting (\ref{eq:s_constraint}) into (\ref{eq:rbm}), we obtain the following \textit{rigid body motion on a sphere} for each rigid body $i\in\{1,\dots,n\}$:\footnote{The property $\hat ab=-\hat ba$ for any 3D vectors $a,b\in\R^3$ is used to obtain the position dynamics (\ref{eq:rbm_sphere_p}).}
\begin{subequations}\label{eq:rbm_sphere}
{\setlength\arraycolsep{2pt}
\begin{eqnarray}
\dot p_{wi}&=&-\rho e^{\hat\xi\theta_{wi}}\hat e_3\omega_i, \label{eq:rbm_sphere_p}\\
\dot e^{\hat\xi\theta_{wi}}&=&e^{\hat\xi\theta_{wi}}\hat\omega_i. \label{eq:rbm_sphere_a}
\end{eqnarray}}
\end{subequations}
\begin{remark}\label{remark:sphere}
{\rm Under the spherical constraint (\ref{eq:s_constraint}), the position of each rigid body $i$ is determined by its attitude $e^{\hat\xi\theta_{wi}}$.
Compared with the rigid body motion (\ref{eq:rbm}), the translational body velocity $v_i$ is also determined by the rotational body velocity $\omega_i$, i.e.,  $v_i=-\rho\hat e_3\omega_i$.
Therefore, the freedom of motion of each rigid body is 3, which is analogous to the 2D vehicle case on a plane\footnote{A plane can be interpreted as the special case of the spherical surface with the radius $\rho=\infty$.} (2D position and 1-dimensional attitude).}% as shown in Fig. \ref{fig:rbm_sphere}.
\end{remark}

From the observation in Remark \ref{remark:sphere}, motion coordination on a sphere, like formation and collision avoidance, is equivalent to attitude coordination on $SO(3)$. 
Therefore, for the convenience of introducing control barrier functions (CBFs) in the subsequent discussion, this letter focuses on attitude control on $SO(3)$ and considers the rotational body velocity $\omega_i$ as the control input of each rigid body $i$.
The actual control input on the sphere is then given by the first two elements of $v_i=-\rho\hat e_3\omega_i=[\rho\omega_{i,y}~-\rho\omega_{i,x}~0]^\T$ and the third element of $\omega_i$ (i.e., $\omega_{i,z}$) for the notation $\omega_i=[\omega_{i,x}~\omega_{i,y}~\omega_{i,z}]^\T$.

%%%%%%%%%%%%%%%%%%%%%%%%%%%%%%%%%%%%%%%%%%%%%%%%%%%%%%%%%%%%%%%%%%%%%%%%%%%%%%%%
\subsection{Rigid Body Network on a Sphere and Research Objective}
In this letter, we suppose that a motion coordination strategy to achieve a control task is given \textit{a priori}, and mainly focus on a distributed collision avoidance problem.
%Therefore, although this section provides the formulation of interconnection topology of the networked rigid bodies for the given motion coordination scheme, this formulation is used later  (presented in Section \ref{subsec:input}).
Here, the interconnection topology between rigid body pairs for the given motion coordination strategy is represented by a directed graph $\G=(\V,\E)$ composed of the rigid body set $\V:=\{1,\dots,n\}$ and edge set $\E\subset \V\times\V$ \cite{book:egerstedt}.
We also define the neighbor set of each rigid body $i$ for the strategy as $\N_i:=\{j\in\V\mid(j,i)\in\E\}$.
Then, $j\in\N_i$ means that rigid body $i$ obtains information about rigid body $j$.

Throughout this work, a group of $n$ rigid bodies with the rigid body motion on a sphere (\ref{eq:rbm_sphere}) and the interconnection topology $\G$ is called a \textit{rigid body network on a sphere}.
This letter has two objectives in this formulation.
The first objective is to develop a new CBF to handle angle-based constraints on $SO(3)$ that also implies motion constraints on a sphere.
The second and main objective is to develop a distributed collision avoidance method for a rigid body network on a sphere based on this kind of CBF.

%%%%%%%%%%%%%%%%%%%%%%%%%%%%%%%%%%%%%%%%%%%%%%%%%%%%%%%%%%%%%%%%%%%%%%%%%%%%%%%%
\section{CONIC CONTROL BARRIER FUNCTIONS}\label{sec:conic}
As a bridge to a collision avoidance problem for a rigid body network on a sphere, this section presents a geometric CBF on $SO(3)$ (refer to \cite{ames_17TAC}-\cite{ames_19ECC} for more details about CBFs). % called a \textit{conic CBF}.
%Then, this concept is extended to a collision avoidance problem for a rigid body network on a sphere in Section \ref{sec:ca}.
We note that only a single rigid body $i$ with the attitude dynamics described in (\ref{eq:rbm_sphere_a}) is considered in this section.

%%%%%%%%%%%%%%%%%%%%%%%%%%%%%%%%%%%%%%%%%%%%%%%%%%%%%%%%%%%%%%%%%%%%%%%%%%%%%%%%
\subsection{Control Barrier Functions on $SO(3)$}
Consider the attitude dynamics (\ref{eq:rbm_sphere_a}), for rigid body $i$ in the world frame $\Sigma_w$, with $e^{\hat\xi\theta_{wi}}\in\cS\subset SO(3)$, $\omega_i\in\Omega\subset\R^3$, and the constraint set $C_o$ defined as
\begin{eqnarray}
C_o:=\{e^{\hat\xi\theta_{wi}}\in SO(3)\mid h(e^{\hat\xi\theta_{wi}})\ge0\}. \label{eq:general_constraint}
\end{eqnarray}
Here, $h: SO(3)\to\R$ is a continuously differentiable function.
Then, similarly to \cite{ames_17TAC}-\cite{ames_19ECC}, we provide a CBF definition on $SO(3)$ as follows:
\begin{definition}\label{definition:ZCBF}
The function $h(e^{\hat\xi\theta_{wi}})$ is called a \textit{zeroing CBF (ZCBF)} defined on the set $\cS$ with $C_o\subseteq\cS\subset SO(3)$, if there exists an extended class $\K$ function $\alpha:\R\to\R$ satisfying
\begin{eqnarray}
\sup_{\omega_i\in\Omega}(\dot h(e^{\hat\xi\theta_{wi}})+\alpha(h(e^{\hat\xi\theta_{wi}})))\ge0~\forall e^{\hat\xi\theta_{wi}}\in\cS. \nonumber
\end{eqnarray}
\end{definition}
%
%
%Notice in Definition \ref{definition:ZCBF} that the control input $\omega_i$ appears in $\dot h(e^{\hat\xi\theta_{wi}})$ from (\ref{eq:rbm_sphere_a}).
%Then, we have the following theorem:
%%
%\begin{theorem}\label{theorem:general}
%If the function $h(e^{\hat\xi\theta_{wi}})$ is a ZCBF on $\cS$, then any Lipschitz continuous controller $\omega_i:\cS\to\Omega$ satisfying $\dot h(e^{\hat\xi\theta_{wi}})+\alpha(h(e^{\hat\xi\theta_{wi}}))\ge0$ will render the set $C_o$ forward invariant.
%\end{theorem}
%%
%\begin{proof}
%By appropriately considering the subspace $\cS\subset SO(3)$, the attitude dynamics (\ref{eq:rbm_sphere_a}) can be rewritten by local Lipschitz continuous dynamics in vector form (see Appendix).
%Corollary 2 in \cite{ames_17TAC} can be thus applied.
%\end{proof}
%%
%
%Theorem \ref{theorem:general} means that the attitude $e^{\hat\xi\theta_{wi}}$ remains in the set $C_o$ for all time.

%%%%%%%%%%%%%%%%%%%%%%%%%%%%%%%%%%%%%%%%%%%%%%%%%%%%%%%%%%%%%%%%%%%%%%%%%%%%%%%%
\subsection{Conic Control Barrier Functions}
%\subsection{Conic Constraint on $SO(3)$}
%
\begin{figure}[t]
\centering
\includegraphics[width=.7\linewidth]{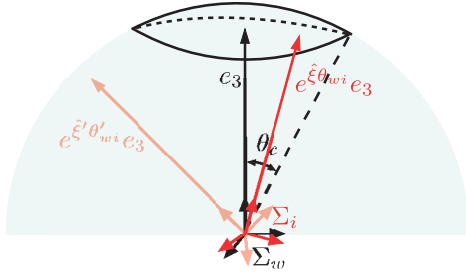}
\caption{Conic constraints. Constraint (\ref{eq:conic_l}): The basis axis $e^{\hat\xi\theta_{wi}}e_3$ is constrained inside and on the boundary of the conic region determined by $e_3$ and $\theta_c$ in $\Sigma_w$. Constraint (\ref{eq:conic_s}): The basis axis $e^{\hat\xi'\theta'_{wi}}e_3$ is constrained outside and on the boundary of the conic region (shown by the semitransparent illustration).}
\label{fig:constraint}
\end{figure}
Let us now provide explicit definitions of $h(e^{\hat\xi\theta_{wi}})$ to represent cone-type (conic) attitude constraints used in this work, which are motivated by \cite{weiss_14ACC}-\cite{nakano_18CDC} and will be extended to collision avoidance techniques in Section \ref{sec:ca}.
Let $e_3$ $(=[0~0~1]^\T)$ be the basis axis in the world frame $\Sigma_w$ and also in the body frame $\Sigma_i$.
Notice then that $e^{\hat\xi\theta_{wi}}e_3$ means the direction of the basis axis of $\Sigma_i$ viewed from $\Sigma_w$.

Consider two kinds of inequality constraints as follows:
\begin{subequations}\label{eq:conic}
\begin{eqnarray}
&&\hspace{-4ex}e^\T_3e^{\hat\xi\theta_{wi}}e_3\ge\cos\theta_c, \label{eq:conic_l} \\
&&\hspace{-4ex}e^\T_3e^{\hat\xi\theta_{wi}}e_3\le\cos\theta_c, \label{eq:conic_s}
\end{eqnarray}
\end{subequations}
where $\theta_c\in(0,\pi/2)$ is the constraint parameter to determine the size of the conic region.
These constraints are formed by the inner product of the basis axes $e_3$ and $e^{\hat\xi\theta_{wi}}e_3$.
The constraint (\ref{eq:conic_l}) (constraint (\ref{eq:conic_s})) thus means that the head of the vector $e^{\hat\xi\theta_{wi}}e_3$ is constrained inside (outside) and on the boundary of the conic region determined by $e_3$ and $\theta_c$ in $\Sigma_w$ (see Fig. \ref{fig:constraint}).
These kinds of constraints are called \textit{conic constraints} in this letter.

%%%%%%%%%%%%%%%%%%%%%%%%%%%%%%%%%%%%%%%%%%%%%%%%%%%%%%%%%%%%%%%%%%%%%%%%%%%%%%%%
%\subsection{Definition of a Conic Control Barrier Function}
We next develop a ZCBF to guarantee the conic constraint (\ref{eq:conic_l}).
Based on (\ref{eq:conic_l}), an angle-based ZCBF, referred to as a \textit{conic CBF} in this work, is defined as $h(e^{\hat\xi\theta_{wi}}):=e^\T_3e^{\hat\xi\theta_{wi}}e_3-\cos\theta_c$.
This enables us to represent the attitude set satisfying the conic constraint (\ref{eq:conic_l}) by (\ref{eq:general_constraint}).

Then, we have the following theorem:
\begin{theorem}\label{theorem:conic}
Any Lipschitz continuous controller $\omega_i:\cS\to\Omega$ satisfying
\begin{eqnarray}
-e^\T_3e^{\hat\xi\theta_{wi}}\hat e_3\omega_i+\alpha(e^\T_3e^{\hat\xi\theta_{wi}}e_3-\cos\theta_c)\ge0 \label{eq:condition}
\end{eqnarray}
will render the set $C_o$ forward invariant.%\footnote{The property $\hat ab=-\hat ba$ for any 3D vectors $a,b\in\R^3$ is used to obtain the condition (\ref{eq:condition}).}% for the attitude dynamics (\ref{eq:rbm_sphere_a}), where $\alpha(\cdot):\R\to\R$ is an extended class $\K$ function.
\end{theorem}
%
%\begin{proof}
\proof
See Appendix. \hfill\QED
%\end{proof}
%
%\begin{proof}
%The condition (\ref{eq:condition}) is directly derived from the condition
%%
%\begin{eqnarray}
%\dot H(e^{\hat\xi\theta_{wi}})+\alpha(H(e^{\hat\xi\theta_{wi}}))\ge0, \nonumber
%\end{eqnarray}
%%
%which means that the function $H$ is a \textit{Zeroing CBF (ZCBF)} \cite{ames_17TAC}.
%Therefore, since the attitude dynamics (\ref{eq:rbm_sphere_a}), given on $SO(3)$, can be rewritten by the vector form satisfying the Lipschitz continuity requirements within the set $C_o$ (see Appendix), Corollary 2 in \cite{ames_17TAC} can be applied.
%\end{proof}
%

Theorem \ref{theorem:conic} means that the attitude $e^{\hat\xi\theta_{wi}}$ remains in the set $C_o$, i.e., inside and on the boundary of the conic constraint (\ref{eq:conic_l}), for all time.
The following corollary also holds for the constraint (\ref{eq:conic_s}):
\begin{corollary}\label{corollary:conic}
Any Lipschitz continuous controller $\omega_i:\cS\to\Omega$ satisfying
\begin{eqnarray}
e^\T_3e^{\hat\xi\theta_{wi}}\hat e_3\omega_i-\alpha(e^\T_3e^{\hat\xi\theta_{wi}}e_3-\cos\theta_c)\ge0 \nonumber %\label{eq:condition2}
\end{eqnarray}
will render the set $C_o$ with $h(e^{\hat\xi\theta_{wi}})=-e^\T_3e^{\hat\xi\theta_{wi}}e_3+\cos\theta_c$ forward invariant.
\end{corollary}
%

%%%%%%%%%%%%%%%%%%%%%%%%%%%%%%%%%%%%%%%%%%%%%%%%%%%%%%%%%%%%%%%%%%%%%%%%%%%%%%%%
\subsection{Safe Control with Conic Control Barrier Functions}
Theorem \ref{theorem:conic} enables us to propose the following attitude control input to guarantee the conic constraint (\ref{eq:conic_l}):
\begin{eqnarray}
\omega^\ast_i=\arg\min_{\omega_i\in\Omega}\|\omega_i-\omega_{nom,i}\|^2~\mathrm{subject~to}~(\ref{eq:condition}). \label{eq:QP}
\end{eqnarray}
Here, $\omega_{nom,i}\in\R^3$ is the nominal controller to achieve a given control task, and the control input is provided by the solution of the quadratic program that can be solved by standard optimization solvers.
The optimization in (\ref{eq:QP}) implies that rigid body $i$ achieves the given control task as much as possible in the sense of minimizing $\|\omega_i-\omega_{nom,i}\|$ while guaranteeing the constraint (\ref{eq:conic_l}).
\begin{figure}[t]
\centering
\includegraphics[width=.75\linewidth]{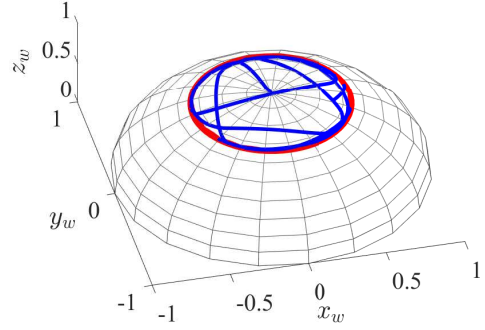}
\caption{Time trajectory of attitude. The blue line shows the time trajectory of the head of the vector $e^{\hat\xi\theta_{wi}}e_3$ in $\Sigma_w$. 
The attitude is constrained inside and on the boundary of the conic constraint (\ref{eq:conic_l}) represented by the red line.}
\label{fig:trajectory}
\end{figure}

As verification, we apply the control input (\ref{eq:QP}) to the attitude dynamics (\ref{eq:rbm_sphere_a}) with $\theta_c=\pi/6$ and a geometric trajectory tracking law as the nominal controller $\omega_{nom,i}$.
Here, the desired trajectory is intentionally set so that the conic constraint (\ref{eq:conic_l}) is violated if the nominal controller is directly applied.
Fig. \ref{fig:trajectory} depicts the time trajectory of the head of the vector $e^{\hat\xi\theta_{wi}}e_3$ in $\Sigma_w$ by the blue line and the boundary of the conic constraint (\ref{eq:conic_l}) by the red one.
This figure shows that the conic constraint (\ref{eq:conic_l}) is satisfied.

%%%%%%%%%%%%%%%%%%%%%%%%%%%%%%%%%%%%%%%%%%%%%%%%%%%%%%%%%%%%%%%%%%%%%%%%%%%%%%%%
\section{COLLISION-FREE MOTION COORDINATION}\label{sec:ca}
\subsection{Collisions on a Sphere}
As stated in Section \ref{subsec:rbms}, this letter focuses on attitude control on $SO(3)$ to deal with motion coordination of a rigid body network on a sphere.
We first define the relative attitude of rigid body $j$ to rigid body $i$ as $e^{\hat\xi\theta_{ij}}:=e^{-\hat\xi\theta_{wi}}e^{\hat\xi\theta_{wj}}\in SO(3)$.
%This relative attitude also represents the relative position under the spherical constraint (\ref{eq:s_constraint}).
Then, we extend the conic CBF approach presented in Section \ref{sec:conic} to a collision avoidance problem by considering the relative attitude case of Corollary \ref{corollary:conic}.

\begin{figure}[t]
\centering
\begin{minipage}[t]{0.285\textwidth}
\centering
\subfigure[Geodesic distance on sphere.]{\includegraphics[width=1\linewidth]{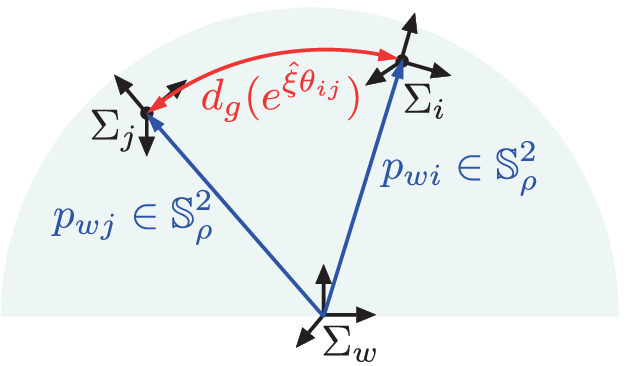}}
\label{fig:g_distance}
\end{minipage}
\hfill
\begin{minipage}[t]{0.175\textwidth}
\centering
\subfigure[Collision.]{\includegraphics[width=1\linewidth]{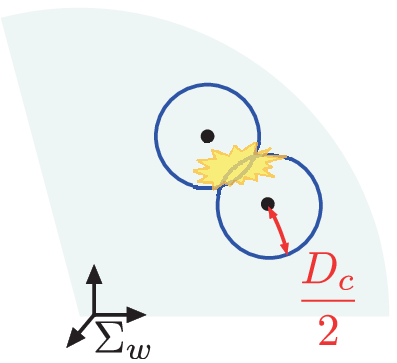}}
\label{fig:collision}
\end{minipage} 
\hfill
\caption{Cartoon illustrations of (a) geodesic distance and (b) collision. (a): The geodesic distance is defined as the arc length of the spherical surface. (b): The collision occurs when the geodesic distance between two rigid bodies is less than the collision distance $D_c$.}
\label{fig:definitions}
\end{figure}

Under the spherical constraint (\ref{eq:s_constraint}), the geodesic distance between rigid body $i$ and rigid body $j$ is defined as the arc length of the spherical surface (see Fig. \ref{fig:definitions}(a)):
\begin{eqnarray}
d_g(p_{wi},p_{wj}):=\rho\cos^{-1}\left(\frac{p^\T_{wi}p_{wj}}{\rho^2}\right). \label{eq:g_distance0}
\end{eqnarray}
Then, substituting (\ref{eq:s_constraint}) into (\ref{eq:g_distance0}) can rewrite $d_g(p_{wi},p_{wj})$ as
\begin{eqnarray}
d_g(e^{\hat\xi\theta_{ij}})=\rho\cos^{-1}(e^\T_3e^{\hat\xi\theta_{ij}}e_3), \label{eq:g_distance}
\end{eqnarray}
that is, the geodesic distance $d_g$ is formed by the relative attitude $e^{\hat\xi\theta_{ij}}$.
\begin{remark}\label{remark:g_distance}
{\rm The geodesic distance $d_g$ is defined by using $\cos^{-1}(\cdot)$, but its argument $e^\T_3e^{\hat\xi\theta_{ij}}e_3$ has a value within the region $[-1,1]$.
Therefore, the geodesic distance $d_g$ is always well defined as the shortest arc length on the spherical surface.
The situation that two rigid bodies exist perfectly at the opposite positions is the special case because we have the infinite number of arcs to determine $d_g$.
However, such an undesired situation can be avoided by appropriately setting distances for collision avoidance (discussed in Section \ref{subsec:ca}).}
\end{remark}

From the geodesic distance definition (\ref{eq:g_distance}), we define collisions between rigid bodies and collision avoidance for a rigid body network on a sphere as follows (see Fig. \ref{fig:definitions}(b)):
\begin{definition}
The collision between rigid body $i$ and rigid body $j$ occurs when $d_g(e^{\hat\xi\theta_{ij}})<D_c$ for the common collision distance $D_c>0$ determined by their shape.
Then, a rigid body network on a sphere is said to achieve \textit{collision avoidance} if
\begin{eqnarray}
d_g(e^{\hat\xi\theta_{ij}}(t))\ge D_c~\forall i,j\in\V~(i\neq j),~t\ge0. \label{eq:ca}
\end{eqnarray}
\end{definition}

In this formulation, we design conic CBFs, derive conditions, and propose a distributed control method for rigid body $i\in\V$ in order to achieve the collision avoidance (\ref{eq:ca}) for a rigid body network on a sphere.
%

%%%%%%%%%%%%%%%%%%%%%%%%%%%%%%%%%%%%%%%%%%%%%%%%%%%%%%%%%%%%%%%%%%%%%%%%%%%%%%%%
\subsection{Distributed Collision Avoidance on a Sphere}\label{subsec:ca}
Define the following safe set $C$ for a rigid body network on a sphere:
\begin{eqnarray}
C:=\{e^{\hat\xi\theta_{wi}},~i\in\V\mid d_g(e^{\hat\xi\theta_{ij}})\ge D_c~\forall i,j\in\V~(i\neq j)\}.\label{eq:safe_set}
\end{eqnarray}
Then, the collision avoidance (\ref{eq:ca}) is equivalent to the forward invariance of the safe set $C$.
Let us now assume that the collision distance $D_c$ satisfies $D_c<(\rho\pi)/2$.
This assumption is reasonable since this inequality means that the geodesic diameter of each rigid body is less than one fourth of the circumference of the sphere, i.e., the size of each rigid body is not too large compared with that of the sphere.

We note that each rigid body is required to take collision avoidance behaviors only when it approaches other rigid bodies.
Besides the graph $\G$ for a given motion coordination strategy, therefore, we introduce another distance-based undirected graph $\G'=(\V,\E'),~\E':=\{(i,j)\in\V~(i\neq j)\mid d_g(e^{\hat\xi\theta_{ij}}) \le D_a~\forall i,j\in\V\}$. % that is often called a \textit{$\Delta$-disk proximity graph} \cite{book:egerstedt}.
Here, $D_a>D_c$ is the geodesic distance within which rigid bodies take account of collision avoidance behaviors.
According to $\G'$, we also define a new neighbor set of rigid body $i$ for the collision avoidance (\ref{eq:ca}), called \textit{distance neighbors}, as $\N_{d,i}:=\{j\in\V\mid(j,i)\in\E'\}$.
Notice now that by employing the reasonable assumption $D_a<(\rho\pi)/2$, we can avoid the undesired special case stated in Remark \ref{remark:g_distance}, i.e., the existence of a distance neighbor perfectly at the opposite position on the sphere, in the collision avoidance process.

Let us define the conic CBF candidates as 
\begin{eqnarray}
h_{ij}:=-e^\T_3e^{\hat\xi\theta_{ij}}e_3+\cos\left(\frac{D_c}{\rho}\right)\in\R,~i,j\in\V, \nonumber
\end{eqnarray}
where we take $\cos(\cdot)$ for the geodesic distance.
Then, by rewriting (\ref{eq:safe_set}) as 
\begin{eqnarray}
C=\{e^{\hat\xi\theta_{wi}},~i\in\V\mid h_{ij}\ge0~\forall i,j\in\V~(i\neq j)\}, \nonumber
\end{eqnarray}
the forward invariance of the safe set $C$ for a rigid body network on a sphere is analogous to collision-free motions.
We now have the following theorem showing the achievement of the collision avoidance (\ref{eq:ca}):
\begin{theorem}\label{theorem:ca}
Suppose that collisions do not occur in a rigid body network on a sphere at the initial time, i.e., $\{e^{\hat\xi\theta_{wi}}(0)\}_{i\in\V}\in C$.
Then, any Lipschitz continuous controllers $\omega_i,~i\in\V$ satisfying
{\setlength\arraycolsep{2pt}
\begin{eqnarray}
e^\T_3e^{-\hat\xi\theta_{ij}}\hat e_3\omega_i&\ge& k\left(e^\T_3e^{\hat\xi\theta_{ij}}e_3-\cos\left(\frac{D_c}{\rho}\right)\right) \nonumber \\
&&\hspace{14ex}\forall j\in\N_{d,i},~k>0 \label{eq:ca_condition}
\end{eqnarray}}
\hspace{-.7ex}will render the safe set $C$ forward invariant.
\end{theorem}
%
%\begin{proof}
\proof
The following condition is first derived from $\dot h_{ij}(e^{\hat\xi\theta_{ij}})+2kh_{ij}(e^{\hat\xi\theta_{ij}})\ge0$ for each rigid body $i\in\V$:
\begin{eqnarray}
&&\hspace{-8ex}e^\T_3e^{-\hat\xi\theta_{ij}}\hat e_3\omega_i+e^\T_3e^{\hat\xi\theta_{ij}}\hat e_3\omega_j \nonumber \\
&&\ge2k\left(e^\T_3 e^{\hat\xi\theta_{ij}}e_3-\cos\left(\frac{D_c}{\rho}\right)\right)~\forall j\in\N_{d,i}. \label{eq:proof}
\end{eqnarray}
Here, $\alpha(h)=2kh,~k>0$ is employed as an extended class $\K$ function, %\footnote{Any extended class $\K$ function can be employed as $\alpha(h)$, e.g., $\alpha(h)=\gamma h^3,~\gamma>0$.
%It is usually selected in view of control requirements and the existence of solutions of the resulting QPs.}
and only the distance neighbors $j\in\N_{d,i}$ are considered because we have $h_{ij}(e^{\hat\xi\theta_{ij}})>0$ for any $j\in\V\setminus\N_{d,i}~(j\neq i)$ from $D_a>D_c$.

The condition (\ref{eq:proof}) for each rigid body $i$ is not distributed since it requires input information of distance neighbors, i.e., $\omega_j,~j\in\N_{d,i}$.
We thus employ the distributed condition (\ref{eq:ca_condition}) to satisfy (\ref{eq:proof}).
Then, because $j\in\N_{d,i}\Leftrightarrow i\in\N_{d,j}$ and $e^\T_3e^{\hat\xi\theta_{ij}}e_3=e^\T_3e^{\hat\xi\theta_{ji}}e_3$ hold, the satisfaction of (\ref{eq:ca_condition}) for all $i\in\V$ guarantees (\ref{eq:proof}) for all $i\in\V$.
Here, considering (\ref{eq:ca_condition}) can be regarded as sharing (\ref{eq:proof}) equally\footnote{As generalization of the equally sharing, we can also introduce weights $w_{ij}\in\R,~(j,i)\in\E'$ satisfying $w_{ij}+w_{ji}=1$ to share the condition (\ref{eq:proof}).} between rigid body $i$ and rigid body $j$.
Corollary \ref{corollary:conic} can be thus applied.\hfill\QED
%\end{proof}
%
\begin{remark}
{\rm The condition (\ref{eq:ca_condition}) is based only on information about distance neighbors $j\in\N_{d,i}$, i.e., distributed.
Moreover, (\ref{eq:ca_condition}) is based only on relative attitude information $e^{\hat\xi\theta_{ij}}$ viewed from $\Sigma_i$ since  $e^{-\hat\xi\theta_{ij}}=(e^{\hat\xi\theta_{ij}})^\T$ holds.}
\end{remark}
%

%%%%%%%%%%%%%%%%%%%%%%%%%%%%%%%%%%%%%%%%%%%%%%%%%%%%%%%%%%%%%%%%%%%%%%%%%%%%%%%%
\subsection{Collision-Free Motion Coordination on a Sphere}\label{subsec:input}
Based on Theorem \ref{theorem:ca}, we propose the following collision-free control input for each rigid body $i\in\V$ in a rigid body network on a sphere:
\begin{eqnarray}
\omega^\ast_i=\arg\min_{\omega_i\in\R^3}\|\omega_i-\omega_{nom,i}\|^2~\mathrm{subject~to}~(\ref{eq:ca_condition}). \label{eq:input}
\end{eqnarray}
Here, $\omega_{nom,i}\in\R^3,~i\in\V$ are the nominal control inputs to achieve a given motion coordination strategy, and each control input is provided by the solution of the distributed quadratic program.
The optimization in (\ref{eq:input}) implies that each rigid body achieves the given motion coordination task as much as possible in the sense of minimizing $\|\omega_i-\omega_{nom,i}\|$ while guaranteeing the collision avoidance (\ref{eq:ca}).
\begin{remark}
{\rm The optimization in (\ref{eq:input}) is always feasible in the safe set $C$ since it has at least one feasible solution $\omega_i=0$.
From the same reason, we can easily impose an additional input saturation constraint, e.g., $\|\omega_i\|\le\omega_{\max}$ for some $\omega_{\max}>0$.
In this case, we can replace $\omega_i\in\R^3$ in (\ref{eq:input}) with $\omega_i\in\Omega,~\Omega=\{\omega\in\R^3\mid\|\omega\|\le\omega_{\max}\}$.}
\end{remark}

Any motion coordination strategy can be applied as $\omega_{nom,i}$ in (\ref{eq:input}).
In the simulation verification presented in Section \ref{sec:simulation}, we apply the following attitude synchronization law \cite{book:hatanaka}:
\begin{eqnarray}
\omega_{nom,i}=k_c\sum_{j\in\N_i}\sk(e^{\hat\xi\theta_{ij}})^\vee. \label{eq:as_input}
\end{eqnarray}
Here, $k_c>0$ is the controller gain, and $\sk(e^{\hat\xi\theta}):=(1/2)(e^{\hat\xi\theta}-e^{-\hat\xi\theta})=\hat\xi\sin\theta\in so(3)$.
Then, it is shown in \cite{book:hatanaka} that if the initial attitudes in a rigid body network satisfy $|\theta_{ij}(0)|<\pi~\forall i,j\in\V$ and the interconnection topology $\G$ is fixed and strongly connected, the control input $\omega_i=\omega_{nom,i}$ given by (\ref{eq:as_input}) achieves the \textit{attitude synchronization} defined as follows:
\begin{eqnarray}
\lim_{t\to\infty}\|e^{\hat\xi\theta_{wi}}(t)-e^{\hat\xi\theta_{wj}}(t)\|_F=0~\forall i,j\in\V. \label{eq:as}
\end{eqnarray}
Here, $\|\cdot\|_F$ is the Frobenius norm.

In the case of a rigid body network on a sphere under the spherical constraint (\ref{eq:s_constraint}), the attitude synchronization (\ref{eq:as}) also implies the \textit{position synchronization} defined as
\begin{eqnarray}
\lim_{t\to\infty}\|p_{wi}(t)-p_{wj}(t)\|=0~\forall i,j\in\V. \nonumber %\label{eq:ps}
\end{eqnarray}
Then, by applying the control input (\ref{eq:input}) with the nominal input (\ref{eq:as_input}), we can expect the achievement of a flocking-like behavior: cohesion; alignment; and separation \cite{reynolds_87CG}, on a sphere.
Here, the final control input of rigid body $i\in\V$ is distributed and based only on relative attitudes with respect to $j\in\N_i\cup\N_{d,i}$, which can be implemented in a distributed manner using onboard sensors, e.g., vision or infrared, without any other communication or global information.

%%%%%%%%%%%%%%%%%%%%%%%%%%%%%%%%%%%%%%%%%%%%%%%%%%%%%%%%%%%%%%%%%%%%%%%%%%%%%%%%
\section{SIMULATION}\label{sec:simulation}
Simulation is carried out to demonstrate the validity of the proposed collision-free motion coordination method (\ref{eq:input}), (\ref{eq:as_input}).
Here, we slightly modify each control input by adding the common rotational body velocity $\omega_c=[0.1~0.2~-0.4]^\T\in\R^3$ to make it easy to see the final configuration of a rigid body network on a sphere.
This modification does not change the relative attitude dynamics, i.e., the same behavior in the sense of the relative states can be seen.

Consider a rigid body network on a sphere with 20 rigid bodies and a strongly connected interconnection topology $\G$.
The simulation parameters are set as $\rho=1$, (i.e., $D_c=\theta_c$), $\theta_c=\pi/150$, $D_a=2D_c$, $k=1$, and $k_c=5$.
The initial positions are set so that each rigid body $i\in\V$ exists in the upper half of the sphere in the $z$-axis direction of $\Sigma_w$.
\begin{figure}[t]
\centering
\subfigure[Position trajectories in $\Sigma_w$.]{\includegraphics[width=.75\linewidth]{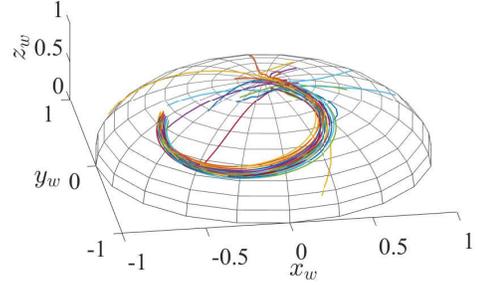}}
\label{fig:trajectory2}
\subfigure[Time response of minimum geodesic distance.]{\includegraphics[width=.72\linewidth]{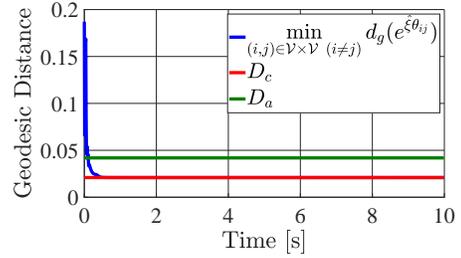}}
\label{fig:distance}
\caption{Simulation results. (a): The rigid body network achieves the cohesion and alignment behaviors on the sphere. (b): The minimum geodesic distance never becomes less than the collision distance $D_c$, i.e., the collision-free motion coordination (separation behavior) is achieved.}
\label{fig:simulation}
\end{figure}

The simulation results are shown in Fig. \ref{fig:simulation}.
Fig. \ref{fig:simulation}(a) illustrates the position trajectories of the rigid body network on a sphere in $\Sigma_w$, which demonstrates the proposed control method achieves the cohesion and alignment behaviors.
The collision avoidance (\ref{eq:ca}) can be confirmed by Fig \ref{fig:simulation}(b) depicting the time response of the minimum geodesic distance between the rigid body pairs $(i,j)\in\V\times\V~(i\neq j)$.

%%%%%%%%%%%%%%%%%%%%%%%%%%%%%%%%%%%%%%%%%%%%%%%%%%%%%%%%%%%%%%%%%%%%%%%%%%%%%%%%
\section{CONCLUSIONS}
This letter presented a distributed collision avoidance control method for a group of multiple rigid bodies on a sphere.
Based on the fact that the rigid body motion constrained to a spherical surface is analogous to the attitude motion on $SO(3)$, the collision avoidance law is derived with conic CBFs on $SO(3)$ that can handle geodesic distance constraints on a spherical surface.
In the proposed method, each rigid body chooses its control input by solving a distributed optimization problem to achieve a given motion coordination strategy while satisfying constraints for the collision avoidance derived by the conic CBFs.
The validity of the proposed approach was demonstrated via simulation.

The future work includes robustness analysis of the proposed conic CBF approaches against disturbances as tackled in \cite{xu_15IFAC,emam_19CDC}.

%The ongoing work is to propose a collision-free cooperative transportation control method for networked multirotor unmanned aerial vehicles.
%Here, the proposed conic CBF plays a key role to avoid inter-collisions since the vehicles have spherical constraints with respect to the payload.


%%%%%%%%%%%%%%%%%%%%%%%%%%%%%%%%%%%%%%%%%%%%%%%%%%%%%%%%%%%%%%%%%%%%%%%%%%%%%%%%
\section*{APPENDIX}
\subsection{Proof of Theorem \ref{theorem:conic}}
\proof
Consider the XYZ (Roll-Pitch-Yaw) Euler angle representation: $(\phi_i,\psi_i,\eta_i)$ to denote the rotation matrix by $e^{\hat\xi\theta_{wi}}=R_x(\phi_i)R_y(\psi_i)R_z(\eta_i)$, where $R_x,R_y,R_z\in SO(3)$ are respectively the basis rotation matrices with respect to $x$-, $y$-, and $z$-axes \cite{book:robomani}.
In the constraint set $C_o$, this rotation matrix can be determined by the Euler parameters with the region $\phi_i,\psi_i\in(-\pi/2,\pi/2)$ since (\ref{eq:conic_l}) becomes $\cos\phi_i\cos\psi_i\ge\cos\theta_c>0$.
Then, with $\zeta_i:=[\phi_i~\psi_i~\eta_i]^\T\in\R^3$, the attitude dynamics (\ref{eq:rbm_sphere_a}) are analogous to the vector form dynamics
\begin{eqnarray}
\hspace{2ex}\dot\zeta_i=\begin{bmatrix}
\frac{\cos\eta_i}{\cos\psi_i} & -\frac{\sin\eta_i}{\cos\psi_i} & 0\\
\sin\eta_i & \cos\eta_i & 0 \\
-\cos\eta_i\tan\psi_i & \sin\eta_i\tan\psi_i & 1
\end{bmatrix}\omega_i=:g(\zeta_i)\omega_i. \nonumber%\label{eq:euler}
\end{eqnarray}

Since $g(\zeta_i)$ consists of smooth trigonometric functions, the attitude dynamics are locally Lipschitz continuous on the subspace $\cS:=\{e^{\hat\xi\theta_{wi}}(\zeta_i)\in SO(3)\mid\phi_i,\psi_i\in(-\pi/2,\pi/2)\}$.
We next consider the boundary of the constraint set $C_o$ denoted by $\partial C_o:=\{e^{\hat\xi\theta_{wi}}(\zeta_i)\in SO(3)\mid h(e^{\hat\xi\theta_{wi}}(\zeta_i))=0\}$.
Then, we obtain $(\partial h)/(\partial\zeta_i)=-\cos\theta_c[\tan\phi_i~\tan\psi_i~0]$ on $\partial C_o$.
This never becomes 0 for $\theta_c\in(0,\pi/2)$ since the attitudes with $\phi_i=\psi_i=0$ are never on the boundary for $\theta_c>0$.
Therefore, Theorem 2 in \cite{ames_19ECC} can be applied. \hfill\QED

%%%%%%%%%%%%%%%%%%%%%%%%%%%%%%%%%%%%%%%%%%%%%%%%%%%%%%%%%%%%%%%%%%%%%%%%%%%%%%%%

\end{document}